\documentclass[aps,prd,twocolumn,nofootinbib,superscriptaddress,floatfix,longbibliography,10pt]{revtex4-2}

\usepackage{amsmath,amssymb,mathtools}
\usepackage{bm}
\usepackage{booktabs}
\usepackage{graphicx}
\usepackage{xcolor}
\usepackage[colorlinks=true,citecolor=blue,urlcolor=blue]{hyperref}
\usepackage{comment}
\usepackage[normalem]{ulem}

\newcommand{\Lag}{\mathcal{L}}

\newcommand{\hc}{\mathrm{H.c.}}
\newcommand{\diag}{\mathrm{diag}}

\newcommand{\LamD}{\Lambda_D}

\begin{document}

\title{Minimal dark $SU(2)$ origin of a massless Dirac neutrino}

\author{P. S. Bhupal Dev}
\affiliation{Department of Physics and McDonnell Center for the Space Sciences, Washington University, Saint Louis, Missouri 63130, USA}
\affiliation{PRISMA$^{++}$ Cluster of Excellence \& Mainz Institute for Theoretical Physics, 
Johannes Gutenberg-Universit\"{a}t Mainz, 55099 Mainz, Germany}

\author{Julia Gehrlein}
\affiliation{Department of Physics, Colorado State University, Fort Collins, Colorado 80523, USA}

\author{Amartya Sengupta}
\affiliation{Department of Physics, The State University of New York, Buffalo, New York 14260, USA}
\affiliation{International Center for Quantum-field Measurement Systems for Studies of the Universe and Particles (QUP),
KEK, 1-1 Oho, Tsukuba, Ibaraki 305-0801, Japan}

\author{Amarjit Soni}
\affiliation{High Energy Theory Group, Physics Department, Brookhaven National Laboratory, Upton, New York 11973, USA}

\begin{abstract}
We propose a gauge-symmetry origin of a rank-two Dirac neutrino mass matrix that enforces one exactly massless neutrino, while being consistent with the oscillation data, as well as cosmological constraints. The mechanism relies on a minimal dark $SU(2)_D$ gauge symmetry under which one right-handed-neutrino-like Weyl fermion is charged, thereby forbidding its Standard Model Yukawa coupling. Quantum consistency then fixes the minimal dark-sector  completion: Cancellation of the Witten anomaly requires a second fermionic $SU(2)_D$
 doublet, while a discrete \(Z_4\) symmetry that forbids Majorana masses allows the two dark doublets to form a vectorlike pair.  This anomaly-free completion gives rise to a secluded, confining dark sector that can contain a potential  dark matter candidate, linking the protected neutrino texture to dark infrared dynamics. 
\end{abstract}

\maketitle

\section{Introduction}
The observation of neutrino oscillations~\cite{Super-Kamiokande:1998kpq,SNO:2002tuh} has increased the number of free parameters in the Standard Model (SM) by at least seven: three mixing angles, three neutrino masses, and at least one CP phase. While 
neutrino oscillation experiments are making steady progress in measuring the oscillation parameters more precisely~\cite{Denton:2025jkt}, with further improvements anticipated from ongoing and next-generation experiments like JUNO~\cite{JUNO:2022mxj}, IceCube/DeepCore~\cite{IceCubeCollaboration:2023wtb}, Hyper-Kamiokande~\cite{Hyper-Kamiokande:2018ofw},  DUNE~\cite{DUNE:2020ypp},
and their combinations~\cite{Nunokawa:2005nx,Fukasawa:2016yue,Agarwalla:2022xdo, Denton:2022een,Parke:2024xre},  
none of the oscillation experiments are sensitive to the absolute neutrino mass scale. The observed solar and atmospheric mass-squared splittings require two non-zero neutrino masses, while the mass of the lightest neutrino cannot be constrained by oscillation experiments, and remains an open question. Instead, we have to rely on cosmological observations of large-scale structure, which constrain the sum of neutrino masses~\cite{DiValentino:2024xsv}, and precise laboratory measurements of the endpoint of the beta-decay spectrum~\cite{Formaggio:2021nfz} to provide complementary information on the neutrino mass scale. The direct laboratory searches, while impressive, are currently far from the sensitivity required to probe even the inverted ordering (IO), let alone normal ordering (NO)~\cite{KATRIN:2024cdt}. Additional information on the neutrino masses can be obtained if a galactic supernova is detected~\cite{Zatsepin:1968kt,Loredo:2001rx,Nardi:2003pr,Pagliaroli:2010ik,Lu:2014zma,Hyper-Kamiokande:2018ofw,Hansen:2019giq,Pompa:2022cxc,Brdar:2022vfr,Pitik:2022jjh,Parker:2023cos,Denton:2024mlb} or a measurement of the cosmic neutrino background is achieved~\cite{PTOLEMY:2019hkd,Alvey:2021xmq}.

Currently, the strongest constraints on the neutrino mass scale come from the 
recent DESI BAO analysis~\cite{DESI:2025zgx}. When interpreted within the \(\Lambda\)CDM cosmology~\cite{Planck:2018vyg,eBOSS:2020yzd}, it   places an upper  
bound on the sum of neutrino masses $\sum_i m_i<64$ meV (95\% CL). Slightly weaker constraints are obtained from Planck~\cite{Planck:2018vyg},  ACT~\cite{AtacamaCosmologyTelescope:2025nti} and SPT~\cite{SPT-3G:2025bzu} datasets, while a combination of ACT+SPT+Planck gives a slightly stronger bound $\sum_i m_i< 62$ meV (95\% CL)~\cite{ACT:2025qjh}. These upper bounds are close to the minimum value of $\sum m_i$ allowed by oscillation data for NO, and already in tension with the corresponding value for IO. Any emerging tension with NO can be alleviated  if the lightest neutrino is  (nearly) massless.  
Although this conclusion might change depending on the assumed cosmological model~\cite{Couchot:2017pvz,Abdalla:2022yfr,Craig:2024tky,Wang:2024hen,Allali:2024aiv,Yadav:2024duq,Green:2024xbb,Elbers:2024sha,Naredo-Tuero:2024sgf,Jiang:2024viw,Bertolez-Martinez:2024wez,Lynch:2025ine,CosmoVerseNetwork:2025alb,Sailer:2025lxj,Giare:2025ath,Graham:2025dqn, Cozzumbo:2025ewt, Allali:2025yvp, Allali:2025wwi} and the presence of additional neutrino interactions~\cite{Beacom:2004yd,Farzan:2015pca,Escudero:2022gez,Benso:2024qrg,Das:2025asx,Sharma:2026ngx}, the increasing precision of cosmological constraints provide a timely motivation to consider theories in which the lightest neutrino remains exactly massless.\footnote{Additional theoretical motivation for a massless neutrino comes from models with a CPT-symmetric universe   
\cite{Boyle:2018tzc,Boyle:2018rgh}.}

Existing models with one massless neutrino rely on a minimal field content of just two right-handed neutrinos instead of the three needed to generate masses for all SM neutrinos. Popular ones include the minimal scotogenic model~\cite{Tao:1996vb,Ma:2006km,Farzan:2012sa,Escribano:2020iqq}, minimal type-I seesaw~\cite{Minkowski:1977sc,Mohapatra:1979ia,Gell-Mann:1979vob,Yanagida:1979as} constructions with two right-handed-neutrinos~\cite{Smirnov:1993af,Ma:1998zg,King:1999mb,Lavoura:2000ci,King:2002nf,Frampton:2002qc,Raidal:2002xf,Ibarra:2003up,King:2015dvf}, and variants of the type-I seesaw such as inverse seesaw~\cite{Mohapatra:1986bd}-based minimal models~\cite{Malinsky:2009df,Dev:2012sg,Abada:2014vea,Camara:2020efq, Thapa:2023fxu}. These models are phenomenologically viable, although if one tries to gauge the usual \(B-L\) symmetry with only two ordinary right-handed neutrinos, additional chiral states or nonstandard charge assignments are required for anomaly cancellation. This prevents these models 
from featuring an  additional gauged $B-L$ symmetry, which is desirable to protect the neutrino mass, and also  for generating a massive gauge boson after lepton-number breaking associated with Majorana mass terms, or embedding it within grand unified theories where \(B-L\) is part of a bigger gauge group.  
An alternative approach is to postulate a texture of the neutrino mass matrix which leads to only two non-zero eigenvalues~\cite{Frampton:2002yf, Xing:2002ta, Ludl:2014axa, Borgohain:2020csn}. 
However, in this case the zero neutrino mass is in general not protected from higher order corrections, rendering it non-zero at a certain energy scale~\cite{Davidson:2006tg}, although in some models the texture zeros are stable~\cite{Fritzsch:2011qv,Dev:2014dla}.  

Here, we instead show that a massless neutrino can follow from a gauge selection rule. Due to its symmetry origin, the zero neutrino mass is protected from corrections and one neutrino remains massless at all energy scales and all orders in perturbation theory. As a minimal example, we introduce  a $SU(2)_D$ gauge symmetry under which one right-handed-neutrino-like SM singlet Weyl fermion $N_1$ transforms as a doublet.  
Furthermore, quantum consistency  constrains the additional dark fermion content, which can accommodate a dark matter (DM) candidate, as shown below.\footnote{$SU(2)_D$ in the context of Dirac neutrino mass and DM has also been considered in Ref.~\cite{Borah:2022phw, Borah:2022dbw}, where the $SU(2)_D$ gauge symmetry is spontaneously
broken to retain a global $SU(2)$ symmetry. Two Dirac neutrino
masses are generated at one loop from the $SU(2)_D$ dark fermions and scalars. In our mechanism, the $SU(2)_D$ gauge symmetry remains unbroken and Dirac neutrino masses arise at tree level. }

\section{The minimal model}
\label{sec:model}
The minimal choice of the dark gauge group is $SU(2)_D$ which features three massless dark gauge fields which couple to the fermions charged under this group with coupling strength $g_D$. 
To obtain an anomaly-safe dark gauge theory, the global Witten anomaly needs to be prevented which requires the introduction of an even number of $SU(2)_D$ doublets~\cite{Witten:1982fp}. Therefore, we include in addition to $N_1$, another $SU(2)_D$ doublet Weyl-fermion field $\xi$. The SM fields and two right-handed neutrinos $N_2,~N_3$ are singlets under the dark gauge group. All new fermions are singlets under the SM gauge groups. 
Additionally, we impose a discrete $Z_4$  symmetry, acting on the SM fermions and on the new singlet and dark-sector fermions, to keep the neutrino sector Dirac.  The masslessness of one neutrino is enforced by the dark gauge charge, which forbids one Dirac Yukawa column.  However, the dark gauge symmetry alone does not exclude Majorana-type bilinears for the SM-singlet fermions; if present, these terms would turn the model into a Majorana or seesaw-like construction rather than a rank-two Dirac-neutrino model.  The chosen $Z_4$ assignment  allows the charged-lepton and Dirac-neutrino Yukawa terms,  forbids Majorana masses, but  permits the mixed dark bilinear involving $N_1$ and $\xi$. 
A $Z_2$ symmetry acting on the same fermions would not be sufficient, since the Majorana bilinears would be even under it.  The field content and their charges are shown in Table~\ref{tab:fields}.

With these charge assignments, the allowed renormalizable Yukawa interactions with the SM lepton and Higgs sector are
\begin{equation}
\Lag_Y=-(Y_e)_{ij}\overline L_iH e_{Rj}
-(Y_\nu)_{i\alpha}\overline L_i\widetilde H N_\alpha+\hc,
\label{eq:yukawa}
\end{equation}
where \(\widetilde H=i\sigma_2 H^\ast\); $i,j=1,2,3$, and  $\alpha=2,3$.
The Majorana-type operators
\begin{equation}
N_\alpha N_\beta,\qquad
\epsilon_{ab} N_1^a N_1^b,\qquad
\epsilon_{ab}\xi^a\xi^b
\label{eq:maj_forbid}
\end{equation}
carry charge \(2\) modulo four and are therefore absent in the \(Z_4\)-symmetric theory.  Here \(\epsilon_{ab}\), with \(\epsilon_{12}=+1\), denotes the antisymmetric \(SU(2)_D\)-invariant tensor used to contract two dark-doublet indices. The mixed dark bilinear $\epsilon_{ab} N_1^a\xi^b$ has charge \(1+3=4=0\mod 4\), so the vectorlike mass

\begin{equation}
\Lag_{\rm mass}=-m_D\,\epsilon_{ab} N_1^a\xi^b+\hc
\label{eq:mass_dark}
\end{equation}
is allowed and it controls the dark infrared dynamics which feature a confinement scale \(\Lambda_D\). 

For \(m_D\gg\Lambda_D\), the dark doublets decouple and the low-energy theory approaches pure \(SU(2)_D\) Yang--Mills; for \(m_D\lesssim\Lambda_D\), the dark fermions remain part of the confined spectrum.  The corresponding dark-sector phenomenology depends on this hierarchy, as discussed later.

The protected zero column in the neutrino mass matrix is not filled by higher-dimensional operators as long as the \(SU(2)_D\) and \(Z_4\) symmetries remain exact.  Any operator that would generate the forbidden Yukawa coupling \(\overline L_i\widetilde H N_1\) carries an uncontracted \(SU(2)_D\) fundamental index unless it is multiplied by a dark-sector operator transforming as a fundamental of \(SU(2)_D\).  Since the \(SU(2)_D\) symmetry is unbroken, and confinement does not break gauge invariance, only \(SU(2)_D\)-singlet operators can acquire vacuum expectation values.  Thus dark confinement can generate singlet bound states or condensates, but it cannot induce the forbidden Yukawa column.  The \(Z_4\) symmetry separately forbids Majorana-type operators, including higher-dimensional lepton-number-violating terms, so the active Dirac neutrino mass matrix remains rank two.

\begin{table}[t]
\centering
\renewcommand{\arraystretch}{1.00}
\begin{tabular}{c c c c}
\toprule
Field
&
\(SU(3)_c\times SU(2)_L\times U(1)_Y\)
&
\(SU(2)_D\)
&
\(Z_4\)
\\
\midrule
\(L_i\)
&
\((\mathbf{1},\mathbf{2},-1/2)\)
&
\(\mathbf{1}\)
&
\(1\)
\\
\(e_{Ri}\)
&
\((\mathbf{1},\mathbf{1},-1)\)
&
\(\mathbf{1}\)
&
\(1\)
\\
\(H\)
&
\((\mathbf{1},\mathbf{2},1/2)\)
&
\(\mathbf{1}\)
&
\(0\)
\\
\(N_1\)
&
\((\mathbf{1},\mathbf{1},0)\)
&
\(\mathbf{2}\)
&
\(1\)
\\
\(\xi\)
&
\((\mathbf{1},\mathbf{1},0)\)
&
\(\mathbf{2}\)
&
\(3\)
\\
\(N_2,N_3\)
&
\((\mathbf{1},\mathbf{1},0)\)
&
\(\mathbf{1}\)
&
\(1\)
\\
\bottomrule
\end{tabular}
\caption{SM lepton and   Higgs, and dark-sector fermion charges under $SU(2)_D$ and \(Z_4\).   
}
\label{tab:fields}
\end{table}
\section{Minimality and alternative choices of the dark symmetry}
\label{sec:minimality}
The zero-column mechanism itself does not require a non-Abelian group; an Abelian dark charge could also forbid \(\overline L_i\widetilde H N_1\). We choose a non-Abelian realization because it also produces a secluded confining sector with an infrared scale generated by strong dynamics. With
this added requirement, \(SU(2)_D\) is the smallest possible choice.  Its minimal consistency issue is the Witten anomaly, already removed by the second doublet \(\xi\) introduced above.

A hidden \(SU(3)_D\) theory is a natural alternative.  It is similar to QCD and can support dark baryons.  But a single Weyl fundamental of \(SU(3)_D\) is perturbatively anomalous. The smallest anomaly-free version introduces an antifundamental partner,\, $\psi_D\sim\mathbf3,\, \eta_D\sim\overline{\mathbf3},
\label{eq:su3_pair}$
making the dark sector vectorlike. Such models may be preferable for dark baryon stability or first-order confinement, but they are less minimal to predict a massless neutrino. Other non-Abelian groups like $SO(N\geq 4)$ and $Sp(N\geq 2)$ can also 
realize the zero-column mechanism, but their spectra and accidental symmetries differ from the minimal \(SU(2)_D\) case.
\section{Neutrino mass matrix}
After electroweak symmetry breaking, the Lagrangian~\eqref{eq:yukawa} generates a Dirac neutrino mass matrix 
\begin{equation}
M_\nu=\frac{v}{\sqrt2}Y_\nu,\qquad
Y_\nu=
\begin{pmatrix}
0&y_{12}&y_{13}\\
0&y_{22}&y_{23}\\
0&y_{32}&y_{33}
\end{pmatrix},
\label{eq:zero_column}
\end{equation}
where $v$ is the electroweak vacuum expectation value.
The first column of $Y_\nu$ is identically zero because \(N_1\) has a dark gauge charge which does not allow its coupling to the SM neutrinos.  
We nevertheless display the texture in a \(3\times3\) square form by reserving the first column for the dark-charged state \(N_1\).
Hence $\det M_\nu=0$, and one physical Dirac neutrino mass vanishes. Working in the charged-lepton mass basis, the Dirac mass matrix is diagonalized by
\begin{equation}
U_{\rm PMNS}^\dagger M_\nu U_{R}
=
\diag(m_1,m_2,m_3),
\label{eq:svd}
\end{equation}
where $U_{R}$ is an arbitrary unitary matrix diagonalizing the $N_i$ sector and $U_{\text{PMNS}}$ is the measurable PMNS  mixing matrix~\cite{Pontecorvo:1957cp,Maki:1962mu}.
Furthermore, since \(N_2\) and \(N_3\) carry identical charges, we can perform a right-handed \(U(2)\) rotation among them:\footnote{This freedom of choosing basis in the \((N_2,N_3)\) space follows solely from the identical quantum numbers of \(N_2\) and \(N_3\), and is independent of the particular choice of dark symmetry group.}
\begin{equation}
\begin{pmatrix}
N_2\\N_3
\end{pmatrix}
\rightarrow
R
\begin{pmatrix}
N_2\\N_3
\end{pmatrix},
\qquad
R\in U(2).
\end{equation}
Let \(B\) be the \(3\times2\) matrix formed by the nonzero columns of \(M_\nu\).  If the second row of \(B\) is \((w_1,w_2)\), choose
\begin{equation}
R^\dagger=
\frac{1}{\sqrt{|w_1|^2+|w_2|^2}}
\begin{pmatrix}
w_2&w_1^\ast\\
-w_1&w_2^\ast
\end{pmatrix}.
\label{eq:rotation}
\end{equation}

Then the second-row entry of the rotated two-column block vanishes, and the full matrix can be written as
\begin{equation}
M_\nu=
\begin{pmatrix}
0&m_{12}&m_{13}\\
0&0&m_{23}\\
0&m_{32}&m_{33}
\end{pmatrix}.
\label{eq:D76}
\end{equation}
This is the $D_{76}$ class of four-zero texture Dirac mass matrix~\cite{Borgohain:2020csn}.\footnote{Four-zero texture neutrino masses have been discussed in Ref.~\cite{Morozumi:2019jrm}. An $SO(10)$ GUT embedding of related texture-zero structures has been studied in Ref.~\cite{Dev:2012xn}.} 
In the present model this is a basis representation, not an additional flavor prediction.
Since the two nonzero columns correspond to fields with identical charges, the additional zero can equivalently be placed in another entry of the second or third column by an appropriate right-handed \(U(2)\) basis choice.  These choices correspond to other $D$-type textures with one zero column in the texture-zero classification, 
but only $D_{76}$ in Eq.~\eqref{eq:D76} and the electron-row representative $D_{16}$ (with $m_{12}=0$) are compatible with the current oscillation data for {\it both} NO and IO~\cite{Borgohain:2020csn}. We focus on the flavor completion of \(D_{76}\) because the physical zero \((M_\nu)_{\mu2}=0\) involves \(\mu\)-row PMNS elements which depend on both \(\theta_{23}\) and \(\delta_{\rm CP}\): \(U_{\mu2},U_{\mu3}\) for NO and \(U_{\mu1},U_{\mu2}\) for IO, and can therefore correlate \(\theta_{23}\) with \(\delta_{\rm CP}\) once the right-handed basis is fixed by additional flavor structure~\cite{supp}. We reiterate that the only basis-independent prediction of the minimal \(SU(2)_D\) construction is the protected zero column, and hence, one massless Dirac neutrino; the \(D_{76}\) texture and the resulting \(\theta_{23}\)--\(\delta_{\rm CP}\) correlations are predictions only of a flavor completion that fixes the right-handed basis.

\section{Phenomenology}
\label{sec:phenomenology}
\subsection{Neutrino sector}
\begin{figure}[t]
\centering
\includegraphics[width=0.96\columnwidth]{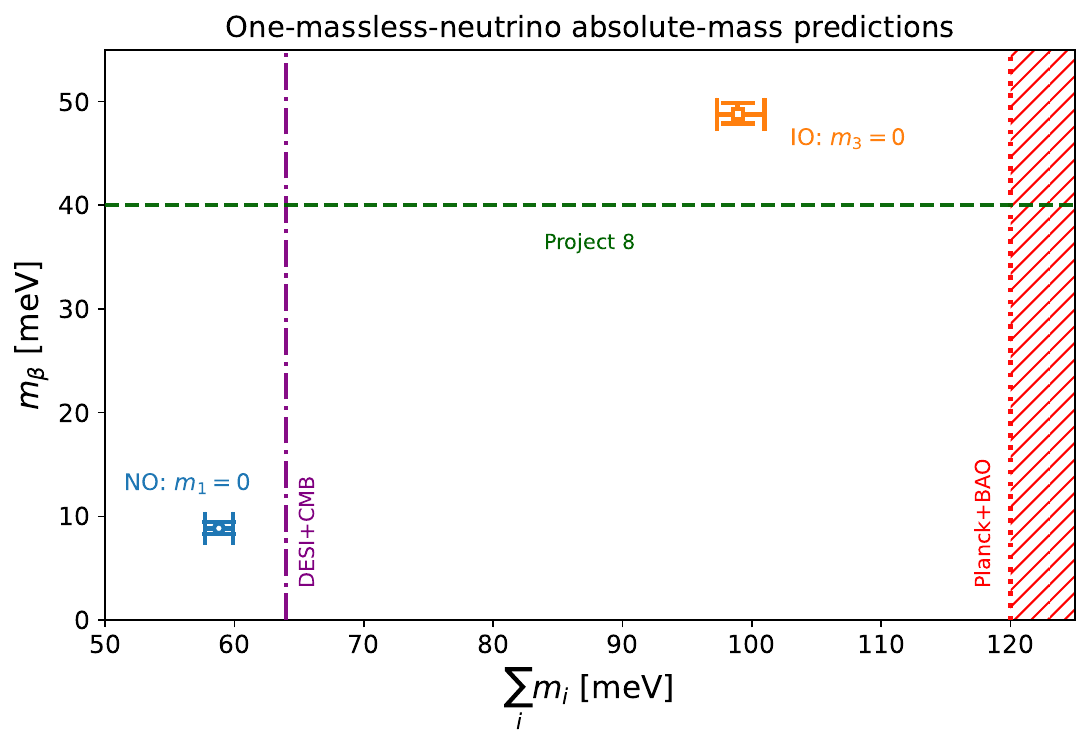}
\caption{
Absolute-mass predictions for the zero-column representative texture with one massless Dirac neutrino.  Error bars show the propagated uncertainties obtained by scanning representative \(3\sigma\) oscillation-parameter ranges in both \(\sum_i m_i\) and \(m_\beta\).  The vertical lines indicate the DESI+CMB~\cite{DESI:2025zgx} and Planck+BAO~\cite{Planck:2018vyg} \(\Lambda\)CDM reference bounds on \(\sum_i m_i\), and the green dashed line shows the approximate Project~8 target sensitivity near \(m_\beta\simeq40\,{\rm meV}\)~\cite{Project8:2022wqh}.  The current KATRIN bound on \(m_\beta\) lies above the plotted range~\cite{KATRIN:2024cdt}.
}\label{fig:mass}
\end{figure}

As stated before, we predict a zero lightest neutrino mass in this model. 
The other neutrino masses are fixed by the measured mass splittings \(\Delta m_{21}^2\) and \(\Delta m_{31}^2\):
\begin{align}
&{\rm NO}:~ m_1=0,\quad m_2=\sqrt{\Delta m_{21}^2},~
m_3=\sqrt{\Delta m_{31}^2}~,
\nonumber 
\\
&{\rm IO:}~ 
m_3=0,~
m_1=\sqrt{|\Delta m_{31}^2|},\quad
m_2=\sqrt{|\Delta m_{31}^2|+\Delta m_{21}^2}. 
\end{align}
The observable related to  the end point of the beta decay spectrum  is~\cite{KATRIN:2024cdt}
\begin{equation}
m_\beta^2=\sum_i |U_{ei}|^2m_i^2~,
\label{eq:mbeta}
\end{equation}
with $U_{ei}$ the PMNS matrix elements in the electron row.
Using the recent JUNO results for the solar sector~\cite{JUNO:2025gmd},
together with the NuFIT~6.0 best-fit values for the remaining oscillation parameters~\cite{Esteban:2024eli}, we find for  Eq.~\eqref{eq:mbeta} and the sum of the neutrino masses:
\begin{align}
&{\rm NO}:~
m_\beta=8.85^{+0.59}_{-0.57}~{\rm meV}, \quad 
\sum_i m_i=58.79^{+1.07}_{-1.04}~{\rm meV}, \nonumber \\
&{\rm IO}:~
m_\beta=48.76^{+1.08}_{-0.88}~{\rm meV}, \quad 
\sum_i m_i=98.92^{+2.06}_{-1.65}~{\rm meV}, 
\end{align}
where the error bars account for the \(3\sigma\) oscillation-parameter uncertainties (see Fig.~\ref{fig:mass}). The predicted values of \(m_\beta\) are far below the current KATRIN bound,
\(m_\beta< 450 ~{\rm meV}\)~\cite{KATRIN:2024cdt}.  
For the sum of neutrino masses,  the NO prediction lies below both the standard Planck+BAO \(\Lambda\)CDM bound \(\sum_i m_i< 120~{\rm meV}\)~\cite{Planck:2018vyg} and the stronger DESI+CMB \(\Lambda\)CDM  bound \(\sum_i m_i< 64~{\rm meV}\)~\cite{DESI:2025zgx}.  The IO prediction   remains allowed by Planck+BAO but disfavored by
DESI+CMB.   

Future improvements in direct beta-decay
searches from experiments like  KATRIN~\cite{Drexlin:2026zam}, ECHo~\cite{ECHo:2025ook}, HOLMES~\cite{Alpert:2025tqq} and  Project~8~\cite{Project8:2022wqh} could eventually reach  sensitivities down to \(m_\beta\simeq40\,{\rm meV}\) and probe the IO case, while the NO prediction is more challenging to reach.  Future CMB and large-scale-structure surveys can further test the mass-sum prediction, with CMB-S4+DESI-like combinations expected to reach \(\mathcal{O}(10\,{\rm meV})\) precision 
in representative forecasts~\cite{Font-Ribera:2013rwa,SimonsObservatory:2018koc,CMB-S4:2016ple,Brinckmann:2018owf}.  This would directly probe the NO prediction near \(59\,{\rm meV}\).  Figure~\ref{fig:mass} summarizes the relevant absolute-mass probes.

This model features Dirac neutrinos which additionally predicts that neutrinoless double beta decay should not be observed. Also, the relic neutrino capture rate is smaller by a factor of two compared to the Majorana case, giving about 4.06 events/year for a 100 g PTOLEMY-like detector~\cite{Long:2014zva}. For a detector with an arbitrarily good energy resolution, each mass eigenstate $\nu_i$ would
make a distinguishable contribution to the C$\nu$B capture and to the beta decay spectrum. The
beta decay spectrum would be the sum of three spectra, and its endpoint would be determined by the
lightest neutrino mass, which offers a way to test whether the lightest mass is zero or not. But in practice, the primary target energy resolution of PTOLEMY is $\sim$50 meV~\cite{PTOLEMY:2019hkd}, so  experimentally, it will be extremely difficult, if not impossible, to determine whether the lightest neutrino is exactly massless or has a nonzero mass smaller than the detector resolution.
\subsection{Dark sector}
\label{sec:Dark_sector}
The neutrino results derived in the previous subsection do not require knowledge of the detailed spectrum of the dark gauge theory;  instead, it  follows  from the non-trivial dark charge of only \(N_1\). 
The dark-sector phenomenology however, depends on the choice of the gauge group and the particle spectrum. 

Once \(SU(2)_D\) is gauged, the theory becomes strong in the infrared and generates its own mass scale by dimensional transmutation~\cite{Gross:1973id,Politzer:1973fx}:
\begin{equation}
\LamD
=
\mu
\exp\left[-\frac{8\pi^2}{b_0g_D^2(\mu)}\right]\, .
\end{equation}
Here \(b_0\) is the one-loop beta-function coefficient.  With \(n_W\) Weyl fermion doublets and \(n_S\) complex scalar doublets of \(SU(2)_D\),
\begin{equation}
b_0=\frac{22}{3}-\frac{n_W}{3}-\frac{n_S}{6}\, .
\end{equation}
In the minimal model \(n_W=2\) and \(n_S=0\), so \(b_0=20/3>0\); the dark gauge theory is therefore asymptotically free and is expected to confine in the infrared.\footnote{Losing asymptotic freedom at one loop would require \(2n_W+n_S\geq44\), which gives \(n_W\geq22\) Weyl doublets in the absence of scalar doublets.}

As in QCD, a gauge theory with no elementary mass scale in the gauge sector can produce massive bound states after confinement.  If the lightest confined states are stable or sufficiently long-lived, the particles in such a secluded confining sector provide natural DM candidates~\cite{Juknevich:2009ji,Hambye:2009fg,Juknevich:2009gg,Bai:2013xga,Boddy:2014yra,Boddy:2014qxa,Yamanaka:2014pva,Buen-Abad:2015ova,Soni:2016gzf,Garani:2021zrr,Forestell:2016qhc,Forestell:2017wov,DeLuca:2018mzn,Berbig:2024uwm}.

The confined spectrum depends on the hierarchy between \(m_D\), the mass of the dark fermions, and the confinement scale \(\Lambda_D\). If $m_D\gg \LamD,$ the dark doublets are heavy compared to the confinement scale, and the low-energy theory is close to pure Yang--Mills.  The lightest states are then dark glueballs. The lightest scalar and pseudoscalar states may be written schematically in terms of the dark field-strength tensor \(G_{\mu\nu}^A\) and its dual \(\widetilde G^{A\mu\nu}\) as
\begin{equation}
S_D\sim G_{\mu\nu}^A G^{A\mu\nu},
\qquad
P_D\sim G_{\mu\nu}^A\widetilde G^{A\mu\nu},
\end{equation}
where \(A=1,2,3\) is the adjoint \(SU(2)_D\) gauge index.   
Their masses are of order
\begin{equation}
m_{S_D},m_{P_D}\sim \kappa\,\LamD,
\end{equation}
where \(\kappa\) denotes nonperturbative coefficients derived from lattice studies of confining Yang--Mills spectra~\cite{Morningstar:1999rf,Lucini:2004my}. If instead
$m_D\lesssim \LamD,$ the dark doublets participate in the confined spectrum.  The theory can then contain fermionic and bosonic dark hadrons built from \(N_1\) and \(\xi\);  for example, states like
\begin{equation}
\epsilon_{ab}N_1^a\xi^b,
\qquad
N_1^\dagger\bar\sigma^\mu\xi 
\end{equation}
arise, where \(\bar\sigma^\mu=(1,-\vec\sigma)\) in two-component Weyl notation and \(\vec\sigma\) are the Pauli matrices.
The detailed spectrum is nonperturbative, but the existence of a secluded confining sector follows directly from the anomaly-safe completion.  

Whether the lightest confined state in our model is a viable DM candidate depends on its stability, portal interactions, relic abundance, and dark-sector thermal history.  In the discussion below, the DM interpretation assumes that the lightest confined state is protected by an accidental dark-sector symmetry or that its decay rate is sufficiently suppressed to make it cosmologically long-lived. In the pure Yang--Mills limit, \(m_D\gg\Lambda_D\), the lightest dark glueball is stable if no portal to the SM is present, and becomes long-lived when its decay proceeds only through higher-dimensional Higgs--dark-glue operators.  If \(m_D\lesssim\Lambda_D\), the lightest dark hadron may instead contain the dark fermions \(N_1,\xi\).  Its stability then depends on the accidental global symmetries of the renormalizable dark-sector Lagrangian.  In particular, the gauge interactions and the vectorlike mass term can preserve a dark fermion number, analogous to baryon number in QCD, even if this symmetry is not imposed explicitly.  Higher-dimensional portal operators or additional ultraviolet (UV) interactions may violate this accidental symmetry and determine the lifetime of the lightest dark hadron.

The leading interactions of the dark sector with the SM arise through higher-dimensional portals.  Since the dark gauge group is non-Abelian, there is no dimension-four kinetic mixing with hypercharge.  The dark gauge theory can, however, contain its own topological term:
\begin{equation}
\Lag_{\theta_D}
=
\frac{\theta_D g_D^2}{32\pi^2}
G_{\mu\nu}^A\widetilde G^{A\mu\nu},
\qquad
\widetilde G^{A\mu\nu}
=
\frac12\epsilon^{\mu\nu\rho\sigma}G^A_{\rho\sigma}.
\label{eq:dark_theta_term}
\end{equation}
This term is internal to the dark sector and is not a portal to the SM.  Its physical meaning depends on the dark-fermion mass structure.  In the chiral limit, when an anomalous axial rotation of the dark fermions is available, \(\theta_D\) can be rotated away. For nonzero \(m_D\), the physical CP-odd parameter is the dark analogue of \(\bar\theta\), schematically
$\bar\theta_D=\theta_D+\arg m_D ,$ up to convention-dependent signs.\footnote{In our model with vectorlike fermions, $\bar\theta_D$ is physical, unlike in $SU(2)_L$ of the SM, where due to the chiral nature of fermions, the electroweak $\theta$-parameter can be rotated away~\cite{Anselm:1993uj,FileviezPerez:2014xju,Brister:2025him}.}  It may affect dark CP violation, the confined spectrum, or the finite-temperature transition, but it does not by itself lead to visible-sector interactions.  The simplest Higgs portals involving dark gauge fields appear at dimension six:
\begin{equation}
\frac{c_G}{\Lambda^2}
(H^\dagger H)G_{\mu\nu}^AG^{A\mu\nu},
\qquad
\frac{\widetilde c_G}{\Lambda^2}
(H^\dagger H)G_{\mu\nu}^A\widetilde G^{A\mu\nu}.
\label{eq:higgs_glue_portals_outlook}
\end{equation}

Compact objects may also constrain parts of the dark-sector parameter space, although the resulting bounds are necessarily model-dependent.  Neutron stars can gravitationally focus DM from their environments ~\cite{Goldman:1989nd,Bertone:2007ae,Garani:2018kkd,Busoni:2021zoe}, but focusing alone does not imply capture: An incoming dark particle must lose energy through scattering, dark self-interactions, or another dissipative process in order to become bound. In the present model such interactions can arise from the Higgs--dark-glue portal operators in Eq.~\eqref{eq:higgs_glue_portals_outlook}, as well as from the confined dark dynamics itself.

If the DM is made of bosonic glueballs and is captured efficiently, its behavior inside the star can differ substantially from that of ordinary nuclear matter.  The characteristic glueball size is set by the confinement scale,
\begin{equation}
R_D\sim \LamD^{-1}
\simeq
2\times10^{-14}\,{\rm cm}
\left(\frac{1\,{\rm GeV}}{\LamD}\right).
\end{equation}
For \(\LamD\sim{\rm GeV}\), the bound states are microscopic and carry internal momenta of order a GeV, larger than the typical neutron Fermi momenta in neutron-star interiors.  Moreover, glueballs are bosons and are not supported by Fermi pressure.  Once captured and thermalized, a sufficiently large population may therefore settle near the stellar center and form a Bose condensate. Efficient annihilation of DM may heat up the neutron star, offering a potential probe in JWST observations of old neutron stars~\cite{Baryakhtar:2017dbj,Garani:2020wge,Ghosh:2026rxi}.

A genuine neutron-star constraint requires the full sequence of capture, thermalization, condensation, and possible gravitational collapse and black-hole growth~\cite{McDermott:2011jp,Bramante:2013hn,Bell:2013xk,Baryakhtar:2017dbj,Dasgupta:2020mqg,Saha:2025fgu}.  For a non-self-interacting boson of mass \(m_{\rm DM}\), the characteristic collapse number scales as
$N_{\rm crit}
\sim M_{\rm Pl}^2/m_{\rm DM}^2$. 
Old neutron stars can therefore exclude part of the parameter space only if capture and thermalization are efficient, and if any black hole formed grows faster than it evaporates.  The relevant rates depend on the portal scale \(\Lambda\), the confinement scale \(\Lambda_D\), and the nonperturbative matrix elements controlling glueball interactions.  In the minimal model \(\Lambda\) is not fixed by the neutrino texture; it parametrizes whatever UV physics generates the higher-dimensional Higgs--glue operators in Eq.~\eqref{eq:higgs_glue_portals_outlook}.  Thus neutron-star bounds can be powerful, but they require a dedicated analysis of the portal strength, capture, thermalization, and collapse conditions.  We leave this dark-sector phenomenology for future work.

Finally, the dark confinement transition  occurs at a temperature of order $T_D\sim \LamD.$ For the minimal \(SU(2)_D\) theory, one should not assume that the transition is strongly first order.  Pure \(SU(2)\) Yang--Mills has a second-order deconfinement transition in the three-dimensional Ising universality class~\cite{Svetitsky:1982gs,Fingberg:1992ju}, and light dark-sector can further change the thermal history, depending on other details such as the reheating temperature.  Other theories, such as larger \(SU(N)_D\) groups or deformed dark sectors, may instead have first-order transitions.  In such cases the same dark gauge framework can produce a stochastic gravitational-wave background~\cite{Schwaller:2015tja,Soni:2016yes,Caprini:2019egz}. Precise predictions on the phase transition and the potential resulting gravitational waves   require a finite-temperature analysis of the dark theory which is left for future work.

\section{Conclusions}
\label{sec:conclusions}  
We have proposed a minimal dark $SU(2)$ model where the lightest neutrino mass is exactly zero and is protected by gauge symmetry.  
We derived the phenomenological predictions in the neutrino sector for this model and showed that it can be probed with future improvements in direct beta-decay searches and cosmological neutrino mass measurements. Additionally, if a flavor symmetry fixes the right-handed neutrino mass basis, the neutrino mass matrix has a four-zero texture  that could also be tested in ongoing and future oscillation experiments.  Beyond the neutrino texture, the same anomaly-safe completion leads to a secluded confining sector with a potential DM candidate, and depending on the model parameters, may provide a promising target for dark sector phenomenology using gravitational-wave and compact-object studies.

\section*{Acknowledgments}
We thank George Parker and Robert Szafron for useful discussions and comments on the draft. We also thank Sharon Adler and Shaouly Bar-Shalom for encouragement that helped bring this work to completion. The work of BD is partly supported by the U.S. Department of Energy under grant No.~DE-SC0017987 and by a Humboldt Fellowship from the Alexander von Humboldt Foundation.  AS is partially supported by the NSF Grant No.~PHY-2310363 and by a QUP fellowship at KEK, Japan.

\appendix
\setcounter{equation}{0}
\renewcommand{\theequation}{A\arabic{equation}}
\setcounter{figure}{0}
\renewcommand{\thefigure}{A\arabic{figure}}
\setcounter{table}{0}
\renewcommand{\thetable}{A\arabic{table}}

\phantomsection
\section*{\texorpdfstring{Supplemental Material:\\[2pt] Flavor completion of the \(D_{76}\) texture}{Supplemental Material: Flavor completion of the D76 texture}}
\addcontentsline{toc}{section}{Supplemental Material: Flavor completion of the D76 texture}
\label{app:dihedral_D76}

We follow the standard non-Hermitian Dirac mass matrix texture-zero notation of Ref.~\cite{Borgohain:2020csn}. In the minimal model, the \(D_{76}\) form is simply a basis representation of the gauge-protected zero-column matrix.  The physical prediction of the minimal construction is one exactly massless neutrino.  Additional flavor structure fixing the \((N_2,N_3)\) basis can make the \(D_{76}\) texture predictive for other observables (see e.g., Refs.~\cite{Feruglio:2019ybq,Denton:2023hkx,Chauhan:2023faf,deMedeirosVarzielas:2025byb}). 

In the diagonal charged-lepton mass basis [cf.~Eq.~\eqref{eq:svd}],
\begin{equation}
M_\nu
=
U_{\rm PMNS}D_\nu U_R^\dagger,
\qquad
D_\nu=\diag(m_1,m_2,m_3),
\label{eq:app_svd}
\end{equation}
so that
\begin{equation}
(M_\nu)_{\alpha\beta}
=
\sum_{i=1}^3
m_i\,U_{\alpha i}\,(U_R)_{\beta i}^{\ast},
\qquad
\alpha=e,\mu,\tau .
\label{eq:app_matrix_element}
\end{equation}
The dark gauge symmetry enforces
\begin{equation}
(M_\nu)_{\alpha1}=0
\qquad
(\alpha=e,\mu,\tau).
\label{eq:app_zero_column}
\end{equation}
Using Eq.~\eqref{eq:app_matrix_element} and the unitarity of \(U_{\rm PMNS}\), this implies
\begin{equation}
m_i(U_R)_{1i}^{\ast}=0
\qquad
(i=1,2,3).
\label{eq:app_UR_alignment}
\end{equation}
Thus the forbidden right-handed direction is aligned with the massless state.  For NO,
\begin{equation}
(U_R)_{1i}=(e^{i\alpha_1},0,0),
\label{eq:app_UR_NO_firstrow}
\end{equation}
whereas for IO,
\begin{equation}
(U_R)_{1i}=(0,0,e^{i\alpha_1}).
\label{eq:app_UR_IO_firstrow}
\end{equation}

Consider the \(D_{76}\) representative texture with the additional zero [cf.~Eq.~\eqref{eq:D76}]
\begin{equation}
(M_\nu)_{\mu2}=0 .
\label{eq:app_D76_extra_zero}
\end{equation}
For NO, the second right-handed direction lies in the massive \((2,3)\) subspace and may be written as
\begin{equation}
(U_R)_{2i}
=
\left(
0,\,
\cos\theta_R,\,
e^{i\phi_R}\sin\theta_R
\right),
\label{eq:app_UR_NO_secondrow}
\end{equation}
up to irrelevant phases.  Here \(\theta_R\) and \(\phi_R\) parameterize the mixing angle and relative phase in the right-handed \((N_2,N_3)\) subspace, respectively.  The zero condition gives
\begin{equation}
m_2U_{\mu2}\cos\theta_R
+
m_3U_{\mu3}e^{-i\phi_R}\sin\theta_R
=0 .
\label{eq:app_mu_zero_NO}
\end{equation}
Using
\begin{equation}
U_{\mu2}
=
c_{12}c_{23}
-
s_{12}s_{23}s_{13}e^{i\delta_{\rm CP}},
\qquad
U_{\mu3}=s_{23}c_{13},
\label{eq:app_Umu_NO}
\end{equation}
where \(s_{ij}\equiv\sin\theta_{ij}\), \(c_{ij}\equiv\cos\theta_{ij}\), and \(\delta_{\rm CP}\) is the Dirac CP phase in the PMNS matrix, the modulus condition gives
\begin{equation}
\tan\theta_R^{\rm NO}
=
\frac{m_2}{m_3}
\frac{|U_{\mu2}|}{|U_{\mu3}|}.
\label{eq:app_thetaR_NO_D76}
\end{equation}
For IO, the second right-handed direction lies in the massive \((1,2)\) subspace,
\begin{equation}
(U_R)_{2i}
=
\left(
\cos\theta_R,\,
e^{i\phi_R}\sin\theta_R,\,
0
\right),
\label{eq:app_UR_IO_secondrow}
\end{equation}
and the zero condition gives
\begin{equation}
m_1U_{\mu1}\cos\theta_R
+
m_2U_{\mu2}e^{-i\phi_R}\sin\theta_R
=0 ,
\label{eq:app_mu_zero_IO}
\end{equation}
with
\begin{equation}
U_{\mu1}
=
-s_{12}c_{23}
-
c_{12}s_{23}s_{13}e^{i\delta_{\rm CP}} .
\label{eq:app_Umu1_IO}
\end{equation}
Taking the modulus gives
\begin{equation}
\tan\theta_R^{\rm IO}
=
\frac{m_1}{m_2}
\frac{|U_{\mu1}|}{|U_{\mu2}|}.
\label{eq:app_thetaR_IO_D76}
\end{equation}
In the minimal zero-column model, Eqs.~\eqref{eq:app_thetaR_NO_D76} and \eqref{eq:app_thetaR_IO_D76} are not constraints.  They become predictions only after a flavor symmetry fixes the right-handed direction.

\begin{figure}[t]
\centering
\includegraphics[width=0.96\columnwidth]{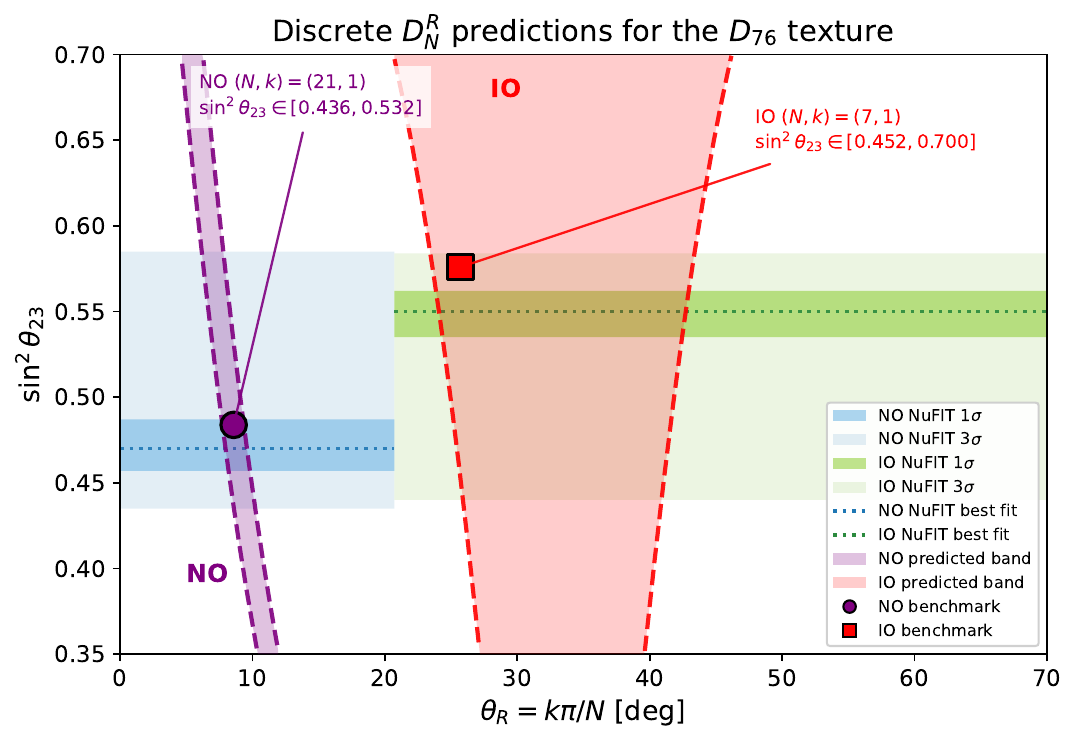}
\caption{Predicted  intervals for the \(D_{76}\) representative four-zero texture Dirac neutrino mass matrix in a right-handed dihedral flavor completion. The right-handed angle is fixed as \(\theta_R=k\pi/N\).  The purple shaded region denotes the NO predicted band, while the red shaded region denotes the IO predicted band; the dashed boundaries show the corresponding edges of the predicted regions after marginalizing over \(\delta_{\rm CP}\).  The blue bands on the NO side show the NuFIT \(1\sigma\) and \(3\sigma\) allowed regions for \(\sin^2\theta_{23}\), while the green bands on the IO side show the corresponding IO NuFIT regions~\cite{Esteban:2024eli}.  The dotted horizontal lines indicate the NO and IO NuFIT best-fit values. The highlighted benchmarks \((N,k)=(21,1)\) for NO and \((N,k)=(7,1)\) for IO are used as representative choices for the fixed flavor+CP correlation plots in Fig.~\ref{fig:D76_theta23_delta}.
}
\label{fig:dihedral}
\end{figure}

As an illustration, we take a right-handed dihedral symmetry \(D_N^R\), where \(D_N\) denotes the symmetry group of a regular \(N\)-gon,\footnote{Some conventions use a group label differing by a factor of two.} with \((N_2,N_3)\) transforming as a doublet.  If \(D_N^R\) is broken to a residual reflection, the direction in the \((N_2,N_3)\) plane is fixed to
\begin{equation}
\theta_R=\frac{k\pi}{N},
\qquad
k=1,\ldots,N-1 .
\label{eq:app_dihedral_angle}
\end{equation}
  For each such discrete value of \(\theta_R\), Eqs.~\eqref{eq:app_thetaR_NO_D76} and \eqref{eq:app_thetaR_IO_D76} select an allowed interval in \(\sin^2\theta_{23}\), after marginalizing over \(\delta_{\rm CP}\), as shown in Fig.~\ref{fig:dihedral}.

For the numerical illustration in Fig.~\ref{fig:dihedral}, we use the NuFIT~6.0~\cite{Esteban:2024eli} oscillation parameter values.  The highlighted choices \((N,k)=(21,1)\) for NO and \((N,k)=(7,1)\) for IO are used below as representative fixed flavor-completion benchmarks.

\begin{figure}[t!]
\centering
\includegraphics[width=0.96\columnwidth]{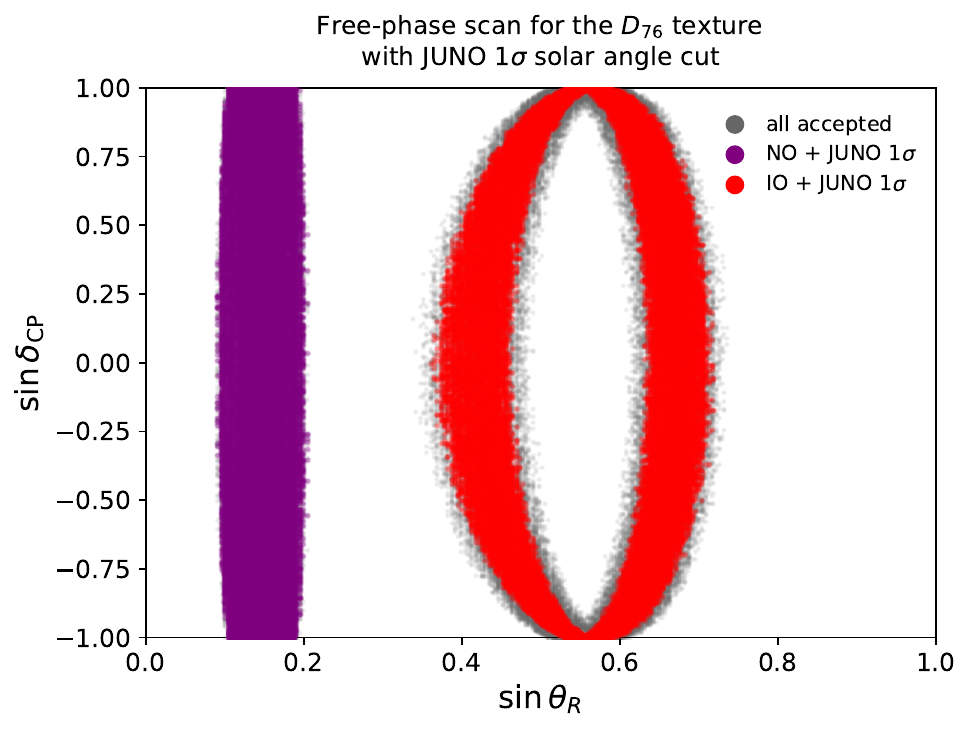}
\caption{
Illustrative right-handed flavor completion with a physical \(\mu\)-row \(D\)-type zero texture (\(D_{76}\)).  The scan is shown in the \((\sin\theta_R,\sin\delta_{\rm CP})\) plane with a free right-handed phase \(\phi_R\).  Gray points satisfy the \(D_{76}\)-type zero condition within numerical tolerance, while purple and red points additionally satisfy the JUNO \(1\sigma\) solar-angle constraint for NO and IO, respectively.
}
\label{fig:mu_D_texture_scan}
\end{figure}

Since \(N_2\) and \(N_3\) have identical gauge and \(Z_4\) charges, a right-handed \(U(2)\) rotation can move the additional zero to another entry of the second or third column without violating any symmetry of the minimal model.  In the minimal theory this remains a basis choice; it becomes physical only after additional flavor structure fixes the \((N_2,N_3)\) basis.  A \(\mu\)-row representative texture then gives a genuine atmospheric-angle--CP-phase correlation.  In Fig.~\ref{fig:mu_D_texture_scan}, we show an illustrative scan with free \(\phi_R\), imposing the zero by matching the magnitudes of the two terms in Eq.~\eqref{eq:app_mu_zero_NO}. Here, we also include the JUNO first-result values for the solar oscillation parameters~\cite{JUNO:2025gmd}
\begin{eqnarray}
\sin^2\theta_{12}&=&0.3092\pm0.0087, \nonumber\\
\Delta m_{21}^2&=&(7.50\pm0.12)\times10^{-5}\,{\rm eV}^2 \, ,
\label{eq:app_JUNO_input}
\end{eqnarray}
together with the remaining oscillation inputs from NuFIT~\cite{Esteban:2024eli} to show the impact of the JUNO results on this texture.

\begin{figure*}[htb!]
\centering
\begin{minipage}{0.485\textwidth}
\centering
\includegraphics[width=\linewidth]{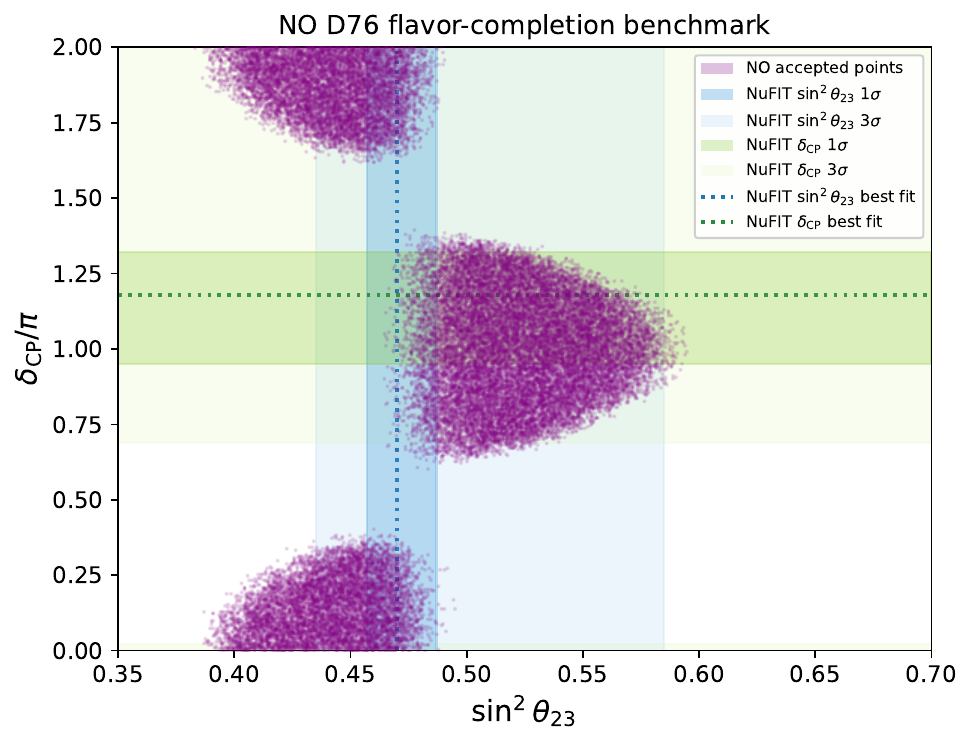}
\end{minipage}
\hfill
\begin{minipage}{0.485\textwidth}
\centering
\includegraphics[width=\linewidth]{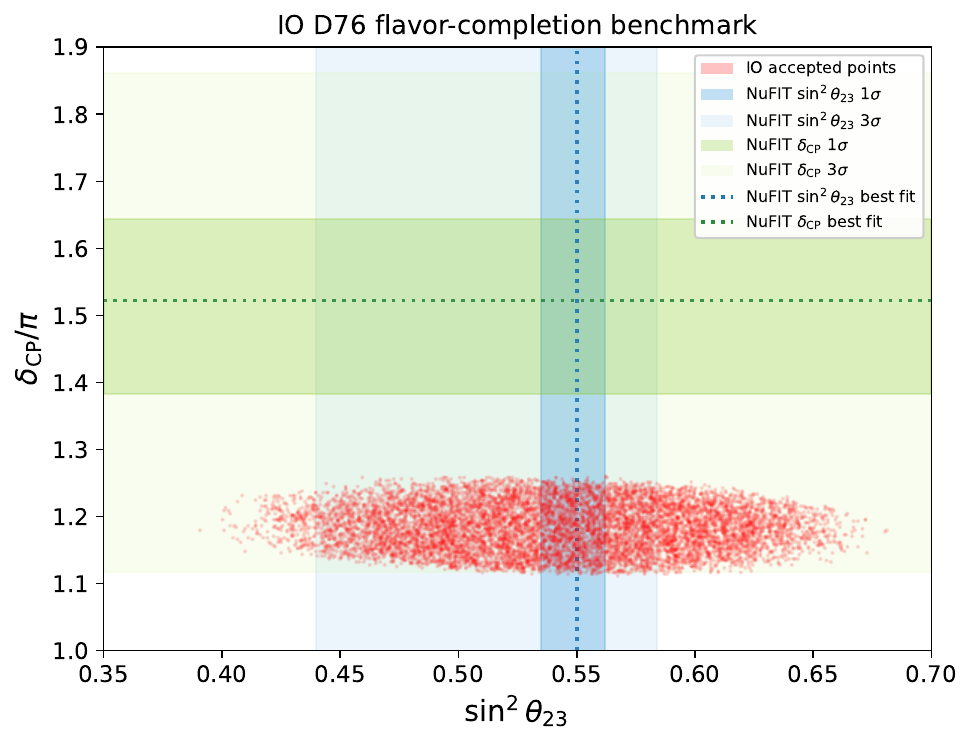}
\end{minipage}
\caption{
Fixed flavor+CP benchmark correlations for the \(D_{76}\) texture in the \((\sin^2\theta_{23},\delta_{\rm CP}/\pi)\) plane.  The left panel shows an illustrative NO benchmark with \(\theta_R=\pi/21\) and \(\phi_R=\pi\), while the right panel shows an illustrative IO benchmark with \(\theta_R=\pi/7\) and \(\phi_R=\pi/14\); see Fig.~\ref{fig:dihedral}.  The points satisfy the \(D_{76}\) zero condition within numerical tolerance together with the JUNO \(1\sigma\) solar-angle cut.  The vertical blue bands and horizontal green bands show the NuFIT \(1\sigma\) and \(3\sigma\) allowed regions for \(\sin^2\theta_{23}\) and \(\delta_{\rm CP}\), respectively~\cite{Esteban:2024eli}.  
The displayed benchmarks are illustrative and not unique; different choices of \((N,k)\) and \(\phi_R\) can lead to different atmospheric-angle--CP-phase correlations.
}
\label{fig:D76_theta23_delta}
\end{figure*}

If the flavor completion fixes not only \(\theta_R\) but also the right-handed phase \(\phi_R\), the \(D_{76}\) condition gives sharper correlations in the \((\sin^2\theta_{23},\delta_{\rm CP})\) plane.  In Fig.~\ref{fig:D76_theta23_delta}, we use the representative choices  selected from Fig.~\ref{fig:dihedral}:  
\begin{eqnarray}
&{\rm NO}:& \theta_R=\frac{\pi}{21},
\qquad
\phi_R=\pi, \nonumber \\
&{\rm IO}:& \theta_R=\frac{\pi}{7},
\qquad
\phi_R=\frac{\pi}{14}.
\end{eqnarray}
The remaining oscillation inputs are scanned over the NuFIT allowed ranges, with the JUNO \(1\sigma\) solar-angle cut imposed. We find that the model prediction for the NO is perfectly within the current NuFIT $1\sigma$-allowed range, while for the representative IO benchmark shown here, the allowed region lies outside the current NuFIT \(1\sigma\) preferred region for \(\delta_{\rm CP}\), but remains within the \(3\sigma\) interval. Of course, these plots are only for illustration purpose; different choices of \((N,k)\) and \(\phi_R\) can lead to different atmospheric-angle--CP-phase correlations.

The flavor-completion discussion in this Appendix is distinct from the absolute-mass constraints in Fig.~\ref{fig:mass} in two ways: (i) Fig.~\ref{fig:mass} is a robust prediction of the  massless lightest neutrino scenario,  independent of any flavor-specific constructions, unlike the correlation plots shown in this Appendix. The dihedral and flavor+CP conditions constrain the right-handed flavor structure through atmospheric-angle--CP-phase correlations, while Fig.~\ref{fig:mass} tests the one-massless-neutrino spectrum through \(m_\beta\) and \(\sum_i m_i\).  (ii) Some IO flavor-completion choices can satisfy the oscillation correlation, although the corresponding one-massless-neutrino spectrum is in tension with the DESI \(\Lambda\)CDM mass-sum bound.
\bibliography{references}

\end{document}